# Estimation of the lifetime of small helium bubbles near tungsten surfaces - a methodological study

Jiechao Cui, Zhangwen Wu, Qing Hou[*]

*Key Lab for Radiation Physics and Technology, Institute of Nuclear Science and Technology, Sichuan University, Chengdu 610061, China*

**Abstract**

Under low energy and high flux/fluence irradiation of helium (He) atoms, the formation and bursting of He bubbles on tungsten (W) surfaces play important roles in the morphological evolution of component surfaces and impurity production in fusion reactors. Microscopically, the bursting of He bubbles is a stochastic process, and He bubbles have statistically average lifetimes. In the present paper, a molecular dynamics-based method was developed to extract, for the first time, the lifetime of He bubbles near tungsten surfaces. It was found that He bubble bursting can be treated as an activated event. Its occurrence frequency or, equivalently, the average lifetime of bubbles follows the Arrhenius equation. For a given bubble size, the activation energy exhibits a good linear dependence with the depth, and the pre-exponential factor obeys the Meyer-Nedle rule. These results are useful for establishing a model in multi-scale simulations of impurity production in the fusion plasma and of the morphological evolution of component surfaces.

[*] Corresponding author. Tel.: +86 28 85412104; fax: +86 28 85410252.

*E-mail address*: qhou@scu.edu.cn (Q. Hou)

# 1. Introduction

As reviewed by a number of authors [1-3], the behavior of materials that are exposed to a complicated and extreme environment in a fusion reactor is a very challenging subject. Among these subjects, the interaction of tungsten (W) with plasma is especially interesting because W has been chosen as a plasma-facing-material (PFM) for the diverter in the International Thermonuclear Experimental Reactor (ITER). In the reactor, the W surface will suffer low energy (<200 eV) and high flux/fluence (~$10^{27}$ m$^{-2}$) irradiation with hydrogen (H) and helium (He) plasma.

Many experiments have been performed to study the He bombardment of W surfaces. The most recent experimental research [4-9] can be mentioned for the low energy and high flux/fluence (LEHF) irradiation of He on W. It is well known from these experiments that, even though the incident energy of He projectiles is significantly lower than the sputtering threshold energy of W, the LEHF irradiation of He can induce severe morphological modifications of W surfaces (e.g., the formation of nano-scale fuzz). Morphological modifications can degrade the physical and chemical properties of surfaces and produce potential impurity sources that would affect the fusion plasma balance. Some either qualitative or phenomenological models have been proposed to explain the experimental observations [4, 10-13]. Although these models were proposed from different viewpoints, it has been commonly recognized that the behavior of He bubbles that are formed in W surfaces plays a central role.

To understand the behavior of He bubbles near W surfaces and to establish the corresponding models, the knowledge of involved atomistic dynamic processes is essential. Atomistic dynamic processes relevant to LEHF irradiation of He on W are plentiful [14], and some of them were addressed by a number of authors based on the molecular dynamics (MD) simulations. For examples, the MD simulations have been performed by Henriksson et al. [15], Lasa et al. [16] and Li et al. [17, 18] to study the growth of He bubbles and W surfaces under the cumulative bombardment of He on W surfaces. Because of the limited time scale that is achievable by the MD simulations, the He fluence in these studies was much lower, and the He bubbles were small compared with those in corresponding experiments. Before carrying forward the simulation investigations to experimental conditions, it is necessary to clarify the contributions of individual processes and their rate information that should be coupled to models of larger time-space scales. The studies of individual processes have been conducted on the reflection of He projectiles on W surfaces [19, 20], interaction of He projectiles with pre-existing He bubbles[21] [22], and on migration and trapping mutation of He atoms and small clusters near W surfaces[23-27]. In this paper, we focus on the bursting of nano-scale He bubbles near W surfaces.

Because the solubility of helium in metals is very low, He bubbles in metal surfaces do not disappear by dissolving into the host metal [28], rather, the bubbles can burst. To extract the critical depth and inner pressure for bursting of He bubbles under palladium surfaces, an MD study , in which the pre-created spherical cavities at a given depth below the surfaces were filled with different number of He atoms and the simulation boxes were then relaxed at one given temperature, was conducted by Wang *et al*[29]. Methodologically similar to the study of Wang *et al* [29], the effects of He/vacancy ratio, temperature and film thickness above the bubbles on the burst of He bubbles near W surfaces were studied by Ito et al. [30]. In the simulation procedure of these two studies, the growth process of He bubbles was omitted. More detailed simulations on the bursting of He bubbles near the W surface were performed by Sefta et al. [31, 32]. The growth process of He bubbles was included by successively adding He atoms to He bubbles every 5 ps. However, the growth rate of the He bubbles so generated was significantly higher than in the experiments. As noted by Sandoval et al. [33], based on the simulations of parallel replica molecular dynamics, the artificial high growth rate may lead to the overestimation of threshold conditions of He bubble bursting. The bursting of He bubbles should occur at a He/Vacancy ratio that is lower than what was predicted by the overrated growth rate of He bubbles. From the MD simulations of cumulative bombardment of He on W surfaces [17], it has been observed that the number of He atoms in He bubbles is approximately twice the number of interstitial W atoms that are produced in the growth of bubbles. Thus, it is most probable that bursting of He bubbles, if it happens, occurs at a He/Vacancy ratio of approximately two. Actually, from the viewpoint of statistical physics, the He bubble bursting is a stochastic process that can eventually happen with a certain probability after the bubbles are formed at a He/Vacancy ratio. The probability is temperature- and size-dependent. Thus, the occurrence frequency of He bubble bursting, or equivalently the average lifetime of He bubbles, should be a more appropriate parameter that can be used to characterize the bursting of He bubbles. Especially for the LEHF irradiation of He on W, the lifetime of He bubbles determines the evolution of bubble density that influences the morphological evolution of surfaces [22].

In the present paper, we propose and evaluate an MD simulation method for calculating the occurrence frequency of He bubble bursting, or equivalently, the average lifetime of He bubbles. The method is a type of emulation of thermal desorption experiments. Although the method had been used by us to calculate the escaping rate of single He atoms trapped near W surfaces [27] and the dissociation rate of W clusters [34], it is the first time to apply the method to the He bubble bursting that consists of more complex atomistic processes. We will show that the triggering of He bubble bursting can be equivalently considered as an activated event of certain activation energy, and the dependence of the occurrence frequency on temperature can be described using the Arrhenius equation.

## 2. Methods

### 2.1. Simulations method

Before running the MD simulation, first, we prepared the initial simulation boxes with the size of $30a_0 \times 30a_0 \times 30a_0$, where $a_0$ represents the lattice constant of W. The surface orientation in the z-direction had a (100) crystal orientations. To create an initial He bubble in the substrate, we created a cavity by removing a given number of W atoms that were closest to a point in the substrate and, then, filled the cavity with helium atoms. The z- coordinate of the point was a given value. The x- and y- coordinates were randomly offset from the substrate center. According to our previous simulations of the He bubble growth in metals [17, 35], the interstitial metal atoms caused by the accumulated pressure of He bubbles move away from the bubbles accompanying the release of the bubble pressure and the restoration of crystal structure surrounding the bubbles. The number of He atoms in the bubbles was found to be approximately twice the number of the ejected metal atoms. Thus, in the present paper, we chose 2 for the ratio $r_{He/v}$ of the number of filling He atoms ($n_{He}$) and the number of the removed W atoms ($n_W$). Fig. 1 shows the initial simulation box, with $d_b$ denoting the thickness of the thinnest ligament above the He bubble, and $R_b$ denoting the initial bubble radius. For each combination of $d_b$, $R_b$, $r_{He/v}$ and surface orientation, we prepared $N_b^{(0)}$ (=100) replicas for the purpose of statistical analysis to be presented in the subsection 2.2.

With the prepared initial simulation boxes, the MD simulations were run using the graphics processing unit (GPU)-based MD package developed in our group [36]. We used the many-body semi-empirical potential of the Finnis-Sinclair type proposed by Ackland et al. for the W-W potential [37]. This W-W potential has been used in our previous work [18, 38, 39] and is widely used by other research groups [26, 31, 33]. For the He-W interaction, we adopted a pairwise potential that was obtained by fitting to the *ab initio* data. The pairwise potential reproduces the correct order of stability for the interstitial He in W [40]. For the He-He interaction, we used the exp-6 potential that was obtained by fitting to the state equation of high pressure He [41].

Throughout all MD runs, the periodic boundary conditions were applied only in the *x*- and *y*- directions. To avoid the drift of substrates during the burst of He bubbles, two bottom layers of the substrates were fixed at their original positions.

Each MD simulation run consisted of three stages. In the first stage, because the initial simulation boxes that were generated above were not in equilibrium, we quenched the boxes to zero temperature. Then, in the second stage, the boxes were, first, thermalized to a temperature $T^{(0)}$ (for example 300 *K*) and, then, relaxed for enough time steps to bring them to thermal equilibrium at this temperature. The thermalization of a simulation box was conducted by assigning the atoms in the box with velocities that were generated by the Monte Carlo sampling of the Maxwell distribution of atom velocity. The boxes were relaxed by numerically integrating the dynamics equation of atoms, using a finite difference scheme that can be written as [42]:

$$\mathbf{r}_i^{(n)} = \mathbf{r}_i^{(n-1)} + \delta t \mathbf{v}_i^{(n-1)} + \frac{1}{2}\delta t^2 \mathbf{a}_i^{(n-1)} \qquad (1.a)$$

$$\mathbf{v}_i^{(n)} = \mathbf{v}_i^{(n-1)} + \frac{1}{2}\delta t(\mathbf{a}_i^{(n-1)} + \mathbf{a}_i^{(n)}) \qquad (1.b)$$

where $\delta t$ is the size of time step. $\mathbf{r}_i^{(n)}$, $\mathbf{v}_i^{(n)}$ and $\mathbf{a}_i^{(n)}$ are the position, velocity and acceleration of *i*th atom at *n*th time step, respectively. Thus far, the simulation procedure is typical. The temperature $T^{(0)}$ is not sufficiently high to cause the He bubble bursting during the simulation time, even if $d_b$ is set to contain two monolayers. To observe the He bubble bursting in the MD simulations, the simulation boxes need to be placed at high temperatures. This was achieved in the third stage.

In the third stage, which corresponds to the analysis method described in the subsection 2.2, we propose a new scheme in which the temperature of simulation boxes is continuously elevated. Assuming the target temperature that the simulation boxes will be raised to in $N_t$ time steps is $T^{(N_t)}$, the temperature increment in one time step is $\Delta T = (T^{(N_t)} - T^{(0)})/N_t$, and the temperature at *n*th time step is:

$$T^{(n)} = T^{(0)} + n\Delta T \tag{2}$$

where $T^{(0)}$ is the initial temperature of boxes of this stage. The finite difference scheme is now written as follows:

$$\mathbf{r}_i^{(n)} = \mathbf{r}_i^{(n-1)} + \delta t \mathbf{v}'^{(n-1)}_i + \frac{1}{2}\delta t^2 \mathbf{a}_i^{(n-1)} \tag{3.a}$$

$$\mathbf{v}_i^{(n)} = \mathbf{v}'^{(n-1)}_i + \frac{1}{2}\delta t(\mathbf{a}_i^{(n-1)} + \mathbf{a}_i^{(n)}) \tag{3.b}$$

$$\mathbf{v}'^{(n)}_i = \mathbf{v}_i^{(n)}\sqrt{T^{(n)}/\tilde{T}^{(n)}} \tag{3.c}$$

where $\mathbf{v}'^{(n)}_i$ represents the scaled velocity of the *i*th atom at the *n*th time step, and $\tilde{T}^{(n)}$ is the instant temperature of simulation boxes calculated by the unscaled velocities, $\mathbf{v}_i^{(n)}$, of atoms. For the convenience of description, we name this scheme a linearly increasing-temperature (LIT) scheme, in contrast to the typical scheme that we denote as the CT scheme in which the ensemble-averaged (or time-averaged) temperature remained constant. Using the LIT scheme, the bursting of He bubbles can be observed as the time progresses. Because the lifetime of He bubbles is concerned, the time point that indicates the starting of He bubble bursting needs to defined. We define this time point when a He atom, which is originally in a bubble, is observed one $a_0$ above the substrate surface. This definition of the He bubble bursting time point is justified by the observation, as to be shown by the results shown, that once a He atom gets out of the substrate, a violent eruption of He atoms immediately occurs.

## 2.2 Extraction of the average He bubble lifetime

From the microscopic view, the bursting of He bubbles is a type of a random event that can be characterized by the occurrence frequency of bursting $f_b(T)$, or the average lifetime $\tau_b(T) \equiv 1/f_b(T)$ of He bubbles. It may have been noted that the LIT scheme of the third stage in our MD simulations is actually an emulation of a thermal desorption experiment, in which the system temperature $T$ is linearly increased with the evolution time. Specifically:

$$T(t) = \beta t + T^{(0)} \qquad (4),$$

where $\beta = (T^{(N_t)} - T^{(0)})/(N_t \delta t)$. Therefore, the analysis method used in thermal desorption experiments can be borrowed to extract $f_b(T)$.

We use $N_b(T)$, with its initial value of $N_b(T_0) = N_b^{(0)}$, to denote the number of simulation boxes in which the He bubbles are intact at the temperature $T(t)$. Because one box contains one He bubble, the rate of He bubble bursting at time $t$ is:

$$\frac{dN_b(T)}{dt} = \beta \frac{dN_b(T)}{dT} = -f_b(T) N_b(T) \qquad (5)$$

If $T_m^{(\beta)}$ is the temperature at which the rate of He bubble bursting is maximum for a give value of $\beta$, we have:

$$\frac{d}{dT}\left(\frac{dN_b(T)}{dT}\right)\bigg|_{T=T_m^{(\beta)}} = -\frac{1}{\beta}\frac{df_b(T)}{dT}\bigg|_{T=T_m^{(\beta)}} N_b\left(T_m^{(\beta)}\right) + \left[\frac{1}{\beta} f_b\left(T_m^{(\beta)}\right)\right]^2 N_b\left(T_m^{(\beta)}\right) = 0$$

$$(6),$$

or in a simplified form:

$$-\frac{df_b(T)}{dT}\bigg|_{T=T_m^{(\beta)}} + \frac{1}{\beta}\left[f_b\left(T_m^{(\beta)}\right)\right]^2 = 0 \qquad (7)$$

If we further assume that the bursting of He bubbles is a type of activated event with a temperature-independent activation energy $E_a$, $f_b(T)$ can be expressed as:

$$f_b(T) = A\, exp\left(-\frac{E_a}{k_B T}\right) \tag{8}$$

From eq.(5) and eq.(6), the following relationship can be derived:

$$2\ln T_m^{(\beta)} - \ln\beta = \frac{E_a}{k_B}\cdot\frac{1}{T_m^{(\beta)}} + \ln\frac{E_a}{k_B A} \tag{9}$$

Using different $N_t$ in our MD simulations for a given $T^{(N_t)}$, a set of $\beta$ values and the corresponding $T_m^{(\beta)}$ can be obtained. The activation energy $E_a$ and the prefactor $A$ can be obtained by fitting eq.(9). On the other hand, the assumption that the bursting of He bubbles is a type of activated event would be justified if eq.(9) can be well-satisfied. An additional point should be mentioned. A small $\beta$ value requires a long computing time to have enough bursting events to extract $T_m^{(\beta)}$, and a large $\beta$ value may cause enough bursting events in a short time. However, the $\beta$ values that are too large may yield results that deviate from the relationship of eq.(9). Thus, the compromise between the computer time and the accuracy of results should be made.

## 3. Results

Based on the abovementioned method, we performed the MD simulation for a number of combinations of ligament thickness $d_b$ and bubble size $n_{He}$ for $r_{He/v} = 2$. The occurrence frequency $f_b(T)$ of bubble bursting [or the average lifetime $\tau_b(T)$] that we are concerned with is assumed to depend on temperature and not on the rate $\beta$ of heating the boxes. This assumption implies that the simulation boxes before the bubble bursting should be in a state of quasi-equilibrium that is independent of the $\beta$ value. Thus, the first point that we observed is the evolution of volume and pressure of He bubbles in the LIT scheme. The instant pressure, $P_b(T(t))$, of He bubbles was calculated using the virial formulation:

$$P_b(T(t)) = \frac{1}{\Omega_b(T(t))} \left( N_{He} k_B T(t) + \sum_{i=1}^{N_{He}} \mathbf{r_i} \cdot \mathbf{f_i} \right) \qquad (8),$$

where $\Omega_b(T(t))$ is the instant volume of He bubbles. Although He bubbles were initially created by filling a spherical cavity of radius $R_b$, the shape of He bubbles at nano-scale is not a well-defined sphere. Thus, we calculate $\Omega_b(T(t))$ by the summation of volumes of He atoms that are defined by the Voronoi polyhedrons of atoms. We implemented an algorithm of the three-dimensional Voronoi tessellation [43] to construct the Voronoi polyhedrons of He atoms. According to the mathematical definition of Voronoi polyhedron, the Voronoi polyhedron of an atom is constructed by the planes that are drawn perpendicular to the vectors joining the atom with its surrounding atoms at the midpoints of the vectors. When the atom moves above the substrate surface (i.e., the He bubble bursting occurs), the construction of its Voronoi polyhedron may fail.

Fig. 2 displays an example of $\Omega_b(T(t))$ and $P_b(T(t))$ against $T(t)$ in the LIT scheme using different $\beta$ values, with $n_{He} = 352$, $d_b = 1.12 a_0$ and the surface orientation in (001) direction. It is observed that different $\beta$ values result in statistically the same $\Omega_b(T(t))$ and $P_b(T(t))$ within the thermal-fluctuation error, with $\Omega_b(T(t))$ and $P_b(T(t))$ only depending on the instant temperature $T(t)$. This result indicates that the simulation boxes in the LIT scheme are in quasi-equilibrium, only if the $\beta$ value is large enough. To validate this assessment further, we also ran MD simulations in the CT scheme at a number of temperatures, in which the simulation boxes were brought to the thermodynamic equilibrium. Corresponding to the case shown in Fig. 2, Fig. 3 displays the volume and pressure of three He bubbles against the time obtained using the CT scheme for $T = 1258.5$. Before the bursting of helium bubbles, the bubble pressures and volumes remained stable around a constant value. The comparison between Fig. 2 and Fig. 3 shows that the bubble volume and bubble pressure in the LIT scheme are consistent with those in the CT scheme when the temperature in the LIT scheme is increased to the same temperature as in the CT scheme. This also suggests that the simulation boxes in the LIT scheme are in quasi-equilibrium. This observation is applicable to other choices of $n_{He}$ and $d_b$ that we considered.

Additionally, from Fig. 2, we observe that the bubble volume linearly expands, and the bubble pressure persists around a constant value against the increasing temperature. The involved atomistic rearrangement of atoms is visualized in Fig. 4, in which the trajectories of W atoms during each time interval are drawn. It is observed that, with increasing temperature, the displacement sequences of W atom along the [111] direction were induced on the upper periphery of He bubbles in association with the volume expansion of He bubbles. The similar phenomenon was observed in the work of Ito et al [30], in which MD simulations were performed at constant temperature. In contrast to what was observed in the work of Sefta et al. [32], where the bubble pressure was accumulated due to the rapid growth of He bubbles until the triggering of loop punching, the He bubble expansion we observe is isobaric. The isobaric expansion is non-isotropic and protrudes at places where the [111] displacement sequences are short. When one He atom penetrates the ligament, the eruption of He atoms immediately occurs as indicated by the sharp peak of the release rate of He atoms shown in Fig. 5. The explosive eruption of He bubbles had been also observed for He bubbles on Pd [29] and Ti [44] surfaces. Thus, the time point at which the first He atom penetrates the ligament can be used to define the beginning of bubble bursting. Using this criteria of bubble bursting, $N_b(T)$, which was defined in subsection 2.2, were obtained. An example of $N_b(T)$ is plotted in Fig. 6 for different $\beta$ values and for $n_{He} = 352$, $r_{He/v} = 2$, $d_b = 1.12a_0$. It can be seen that there is a temperature range in which the number of broken bubbles rapidly increases. However, because only 100 simulation boxes ( $N_b^{(0)} = 100$ ) were used for each $\beta$ value (due to the available computer time), using the differential of $N_b(T)$ to extract $T_m^{(\beta)}$ that is defined by eq.(5) - (7) has large uncertainty. Thus, based on the observation that $N_b(T)$ is approximately symmetric at $N_b(T) = 50$, we used the temperature at $N_b(T) = 50$ to estimate $T_m^{(\beta)}$. We used the same method to estimate $T_m^{(\beta)}$ for all cases considered in the present paper.

Fig. 7a) displays $2\ln T_m^{(\beta)} - \ln\beta$ as a function of $1/T_m^{(\beta)}$ for different ligament thicknesses: $d_b = 0.89a_0$, $1.12a_0$, $1.42a_0$, $1.72a_0$, $1.92a_0$ and $2.22a_0$, and for a given bubble size: $n_{He} = 352$. For $d_b \leq 1.72a_0$, $\beta$=270 K/ps, 135 K/ps, 67.5 K/ps, 38.6 K/ps and 27 K/ps were applied in the LIT scheme. For larger $d_b$, $\beta$=135 K/ps, 67.5 K/ps, 38.6K/ps, 27 K/ps and 13.5 K/ps were applied. It is seen that the data can be well fitted to eq.(9) with fitting parameters $E_a$ and $A$ for all considered $d_b$ values. These results indicate that the bursting of He bubbles for a given $d_b$ can be considered as a type of activated event, and the occurrence frequency of bursting follows the Arrhenius relation, which is an assumption leading to the derivation of eq.(9). The activation energy $E_a$ and the pre-exponential factor $A$ are dependent on $d_b$ and are shown in Table 1. Correspondingly, Fig. 7b) shows the activation energy $E_a$ as a function of $d_b$. Considering the statistical uncertainty, $E_a$ is linearly proportional to $d_b$ and can be fitted to: $E_a(d_b) = a \cdot d_b + b$ with $a$=1.11 eV/$a_0$ and $b$=-0.75 eV. For the activated events, the Meyer-Nedle rule is often followed, which states that the increased activation energy may be compensated by the increased pre-exponential factor for a family of activated processes[45]. According to this rule, the pre-exponential factor $A$ and the activation energy $E_a$ can be empirically related using $\ln A = p \cdot E_a + q$, with $p$ and $q$ being constant. In Fig. 7c), we show $A$ vs $E_a$ for different $d_b$. It is seen that, although the data are scattered, the Meyer-Nedle rule is obeyed, and the relation can be approximately written as $\ln A(E_a) = 3.05 E_a + 0.83$ with $E_a$ in units of eV and $A$ in ps$^{-1}$.

Fig. 8a) displays $2\ln T_m^{(\beta)} - \ln \beta$ as a function of $1/T_m^{(\beta)}$ for different bubble sizes: $n_{He}$= 100, 200, 352 and 486, and given ligament thickness: $d_b = 0.89 a_0$. Again, the data obtained by the MD simulation using the LIT scheme is well fitted by eq.(9). The dependence of $E_a$ on $n_{He}$ is illustrated in Fig. 8b) and can be approximately described by the relation $E_a^{1/3}(n_{He}) = 13.96/n_{He} + 0.5$. However, the relation between $A$ and $E_a$ does not follow well the Meyer-Nedle rule [Fig. 8c)]. The dependence of $E_a$ on $n_{He}$ suggests that, for He bubbles with the same ligament thickness, the small bubbles are more likely to burst than the large bubbles.

In a previous study on helium bubble pressure in bulk W [38], we have shown that a helium bubble in tungsten consists of a core and an interface of finite thickness. The core is compressed more and, thus, the bubble pressure is higher for small bubbles than for large bubbles. This is also true for cases when the bubbles are close to W surfaces, as demonstrated in Fig. 9. It is observed that the pressure, under which the He bubbles with the same initial ligament thickness expand isobarically, is higher for small bubbles than for large bubbles. The results indicate that the bubble interface has the effect that prevents He bubbles from expansion and bursting. It is suggested that the deterministic manner, in which the criteria of bubble bursting is based on the bubble pressure, may be inadequate.

## 4. Concluding remarks

An MD simulation method for estimating the average lifetime of He bubbles has been described. The method is validated for a number of bubble depths and bubble sizes. The results show that the He bubble bursting can be treated as an activated event with its occurrence frequency following the Arrhenius equation. For a given bubble size, the activation energy exhibits good linear dependence on depth, and the pre-exponential factor obeys the Meyer-Nedle rule. These results are informative for establish a model of growth and release of He bubbles, which is closely related with the rate of impurity production in the fusion plasma and the morphological evolution of component surfaces, especially when the lifetimes of He bubbles are comparable with the implantation rate of He atoms. Based on the data obtained in the present paper for activation energy and pre-exponential factor, the scale of average lifetimes of the bubbles can vary from picoseconds to microseconds in a temperature range of 1000K~2000K, depending on the bubble depth and size. This temperature range is interesting for the LEHF experiments of He irradiation of W. Due to the limitation of available computing resources, the emphasis of our paper is on the introduction and evaluation of the method. Certainly, more calculations that cover a larger parameter space are needed to establish a comprehensive knowledge and data base.

In contrast to MD simulations in which He atoms were successively added to the bubbles, He bubble growth is not considered in the present paper, and the number of He atoms in the bubbles was kept constant. It is known that self-interstitial W atoms can be induced around He bubbles during the growth process [31] [46, 47]. The existence of self-interstitial W atoms may change the condition of He bubble bursting. In addition, the time scale that self-interstitial W atoms sustain around He bubbles is a problem that requires study. We will consider this problem in our future work.

## Acknowledgements

This work was supported in part by the National Magnetic Confinement Fusion Program of China (2013GB109002).

**Table. 1.** Activation energy $E_a$ and the preexponential factor $A$ extracted by fitting eq.(9) for the bubble size $n_{He} = 352$ at different depths.

| $d_b(a_0)$ | 0.98 | 1.12 | 1.42 | 1.72 | 1.92 | 2.22 |
|---|---|---|---|---|---|---|
| $E_a$(eV) | 0.198 | 0.532 | 0.744 | 1.35 | 1.304 | 1.683 |
| $A$(ps$^{-1}$) | 1.864 | 13.077 | 20.82 | 190.164 | 35.97 | 59.26 |

**Fig. 1.** Schematic graph of the initial simulation box. The dark spheres are He atoms, the circles are W atoms. The two bottom layers of W atoms are fixed (read circle).

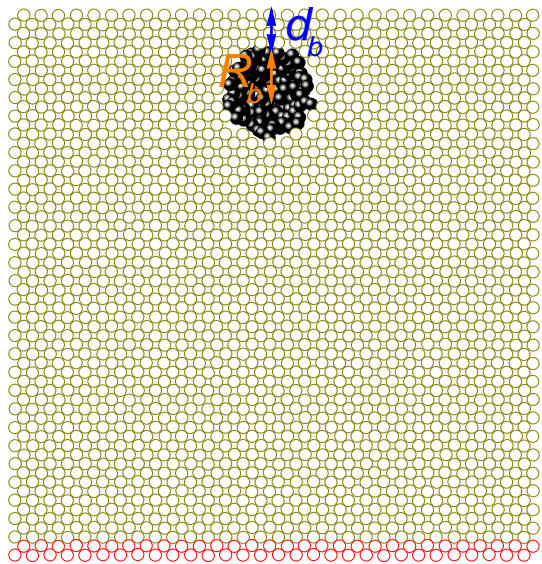

**Fig. 2.** Example of evolution of bubble pressure $P_b(T(t))$ and bubble volume $\Omega_b(T(t))$ vs. temperature evolution $T(t)$ in the LIT scheme with different $\beta$ values and $n_{He} = 352$ and $d_b = 1.12a_0$. The curves end when the bubbles burst. a) Bubble pressure. b) Bubble volume.

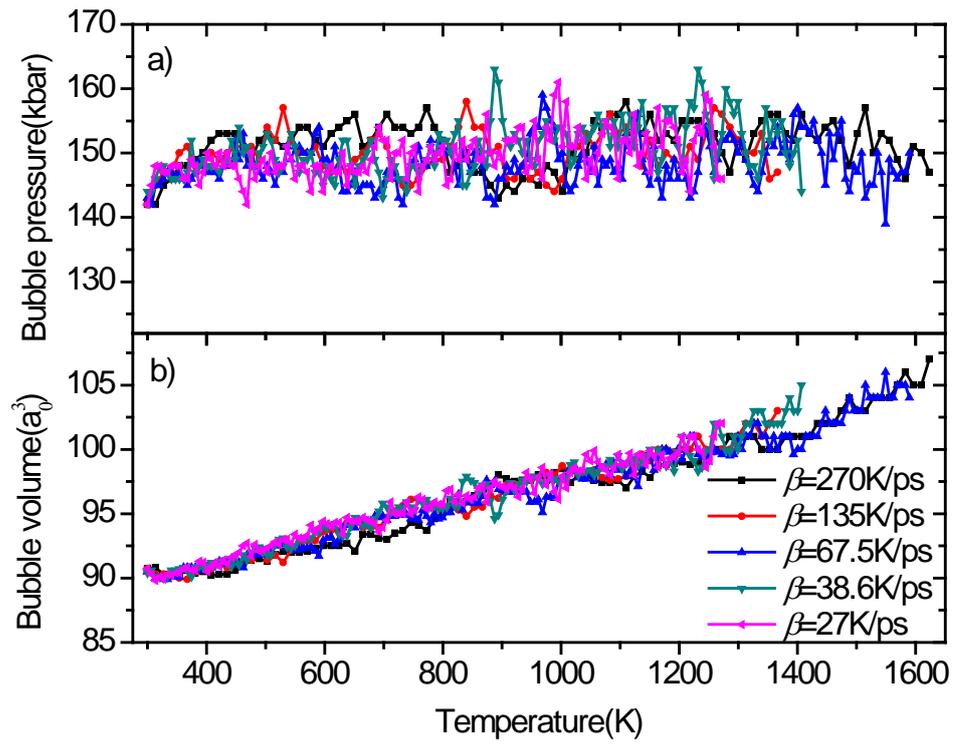

**Fig. 3.** Examples of the instant bubble pressure $P_b(t)$ and bubble volume $\Omega_b(t)$ in the CT scheme at a given temperature of 1258.5 K, with $n_{He} = 352$ and $d_b = 1.12a_0$. The curves end when the bubbles burst. a) Bubble pressure. b) Bubble volume.

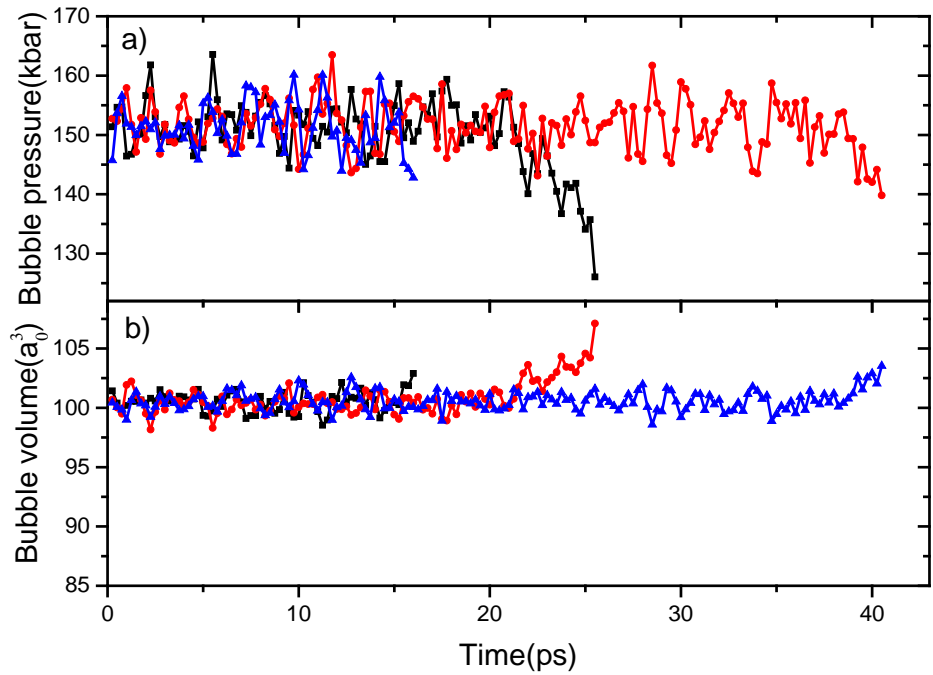

**Fig. 4.** Visualization example of the LIT scheme box configuration at temperature $T$, viewed in x-direction. Circle: He atoms; dot: tungsten atoms. Also drawn (lines) are the trajectories of tungsten atoms when temperature increases from $T1$ to $T$. For a) – c), the temperature interval ($T1$, $T$) = (300K ~ 570K), (1110,K 1380K), (1380K,1650K), and (1650K, 1920K).

a) 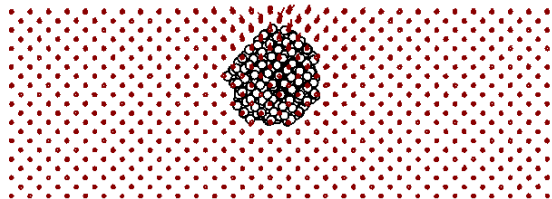

b) 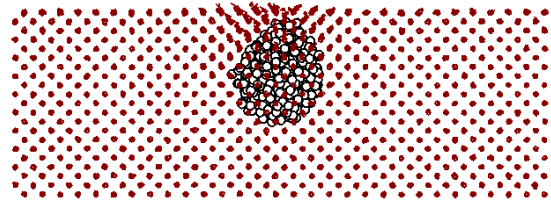

c) 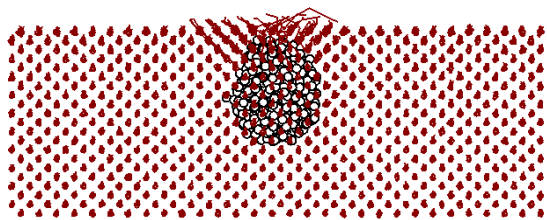

d) 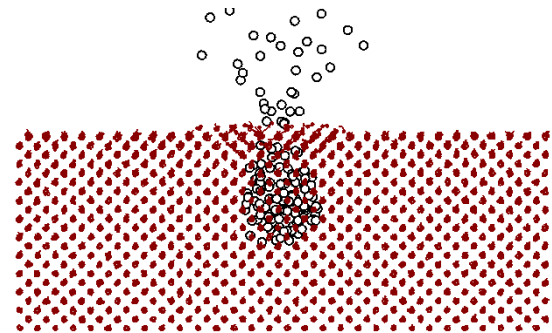

**Fig. 5**. Example of the rate of He atom escaping from the substrate with $n_{He} = 352$ and $d_b = 1.12a_0$.

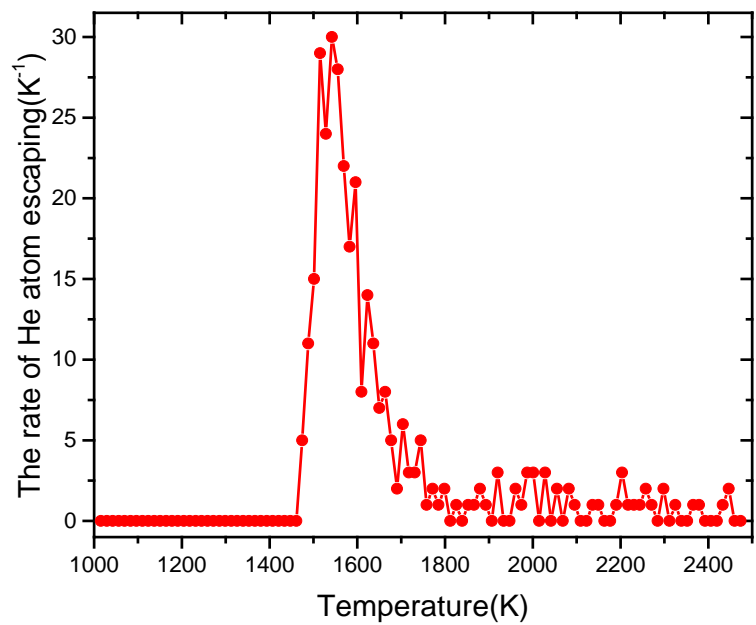

**Fig. 6.** Evolution of $N_b(T(t))$ vs. $T(t)$ obtained in the LIT scheme for different $\beta$ values and $n_{He} = 352$ and $d_b = 1.12a_0$. The dotted lines are drawn to denotes the extracted $T_m^{(\beta)}$.

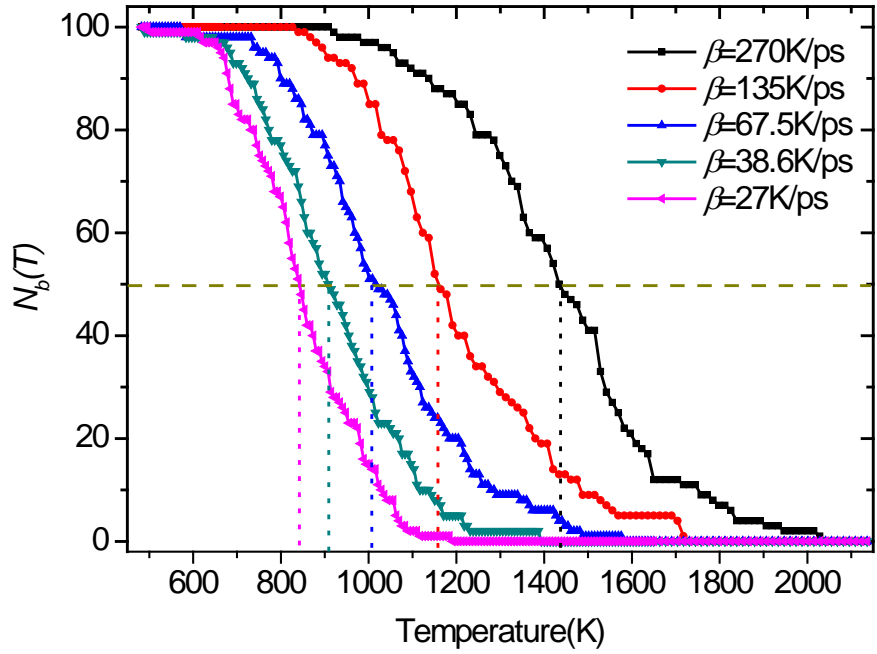

**Fig. 7.** a) $2\ln T_m^{(\beta)} - \ln \beta$ as a function of $1/T_m^{(\beta)}$ [see eq.(9)], obtained for a number of ligament thicknesses $d_b$ with $n_{He} = 352$ and $r_{He/v} = 2$. The lines are the corresponding fittings of eq.(9) with $E_a$ and $A$ being the fitting parameters. b) The activation energy $E_a$ vs. the ligament thicknesses $d_b$ for $n_{He} = 352$ and $r_{He/v} = 2$. The solid line is the fitting function $E_a(d_b) = a \cdot d_b + b$ with $a$=1.11 eV/$\alpha_0$ and $b$=-0.75 eV. c) $\ln A$ vs. $E_a$. The solid line is the fitting function $\ln A(E_a) = 3.05 E_a + 0.83$.

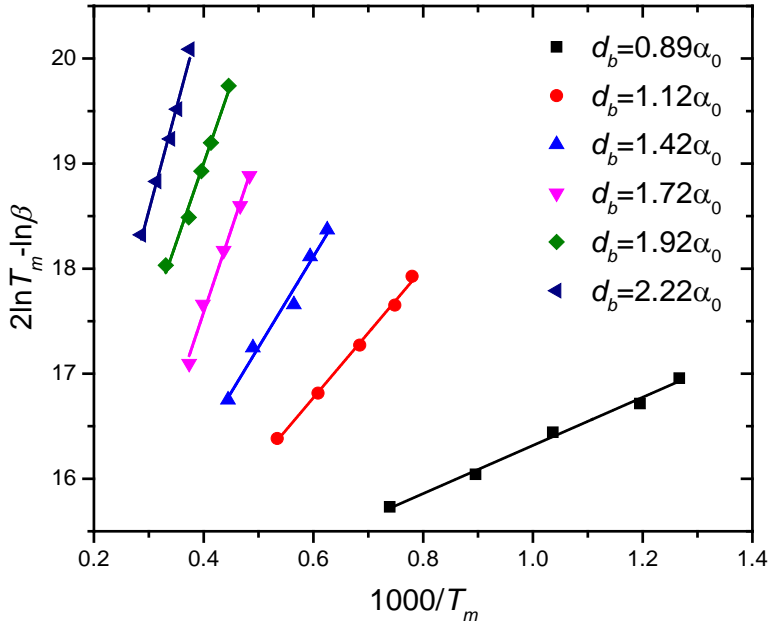

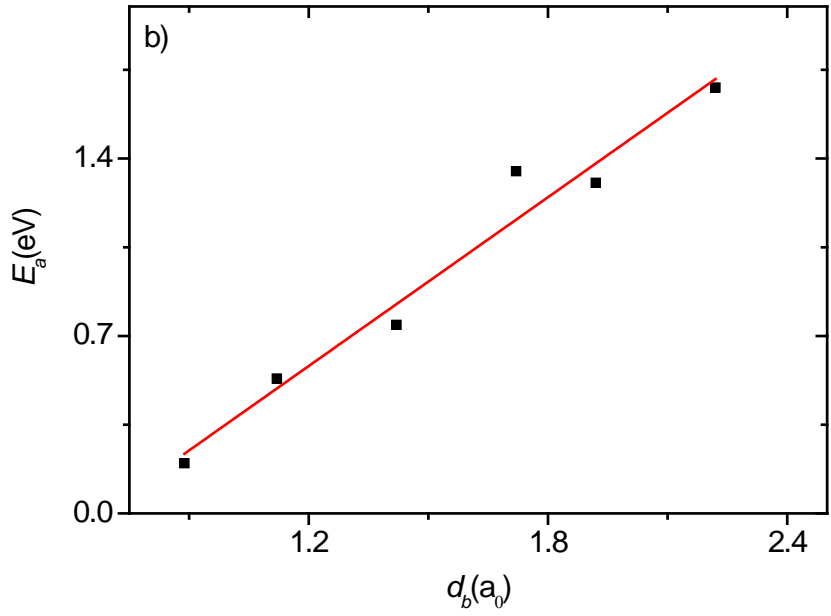

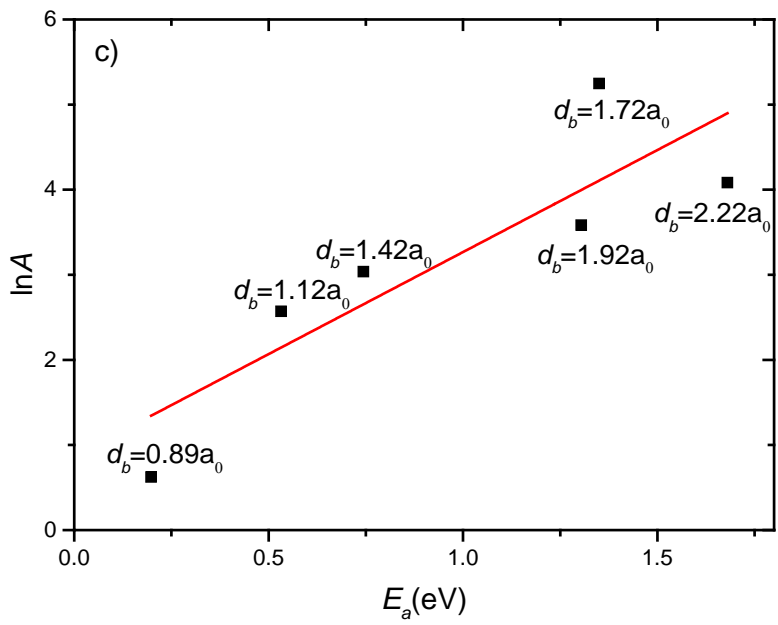

**Fig. 8.** a) $2\ln T_m^{(\beta)} - \ln\beta$ as a function of $1/T_m^{(\beta)}$ [see eq.(9)], obtained for a number of bubble sizes, with $d_b = 0.89\alpha_0$ and $r_{He/v} = 2$. The lines are the corresponding fittings of eq.(9) with $E_a$ and $A$ being the fitting parameters. The lines are the corresponding fittings of eq.(9) with $E_a$ and $A$ being the fitting parameters; b) The dependence of $E_a$ on the bubble size $n_{He}$. The solid line is the fitting function $E_a^{1/3}(n_{He}) = 13.96/n_{He} + 0.5$; c) $\ln A$ vs. $E_a$.

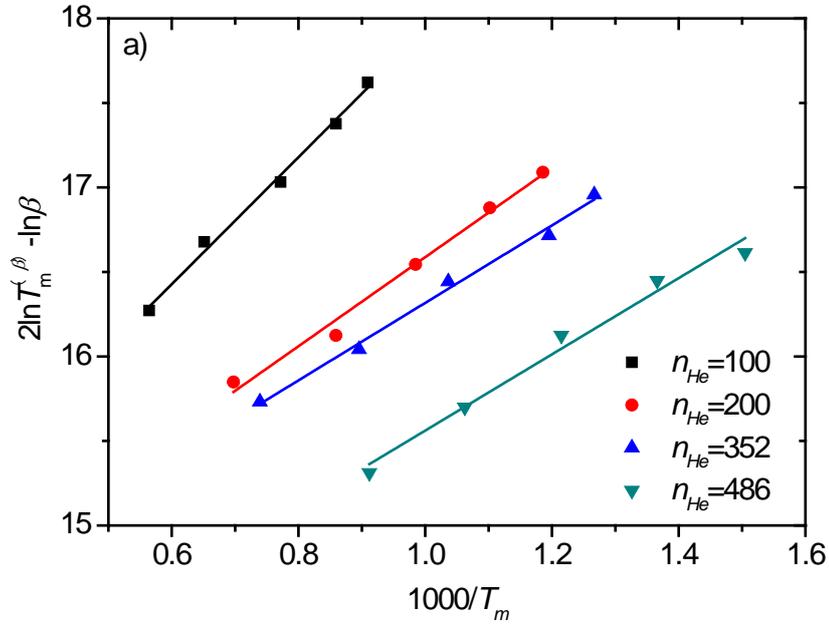

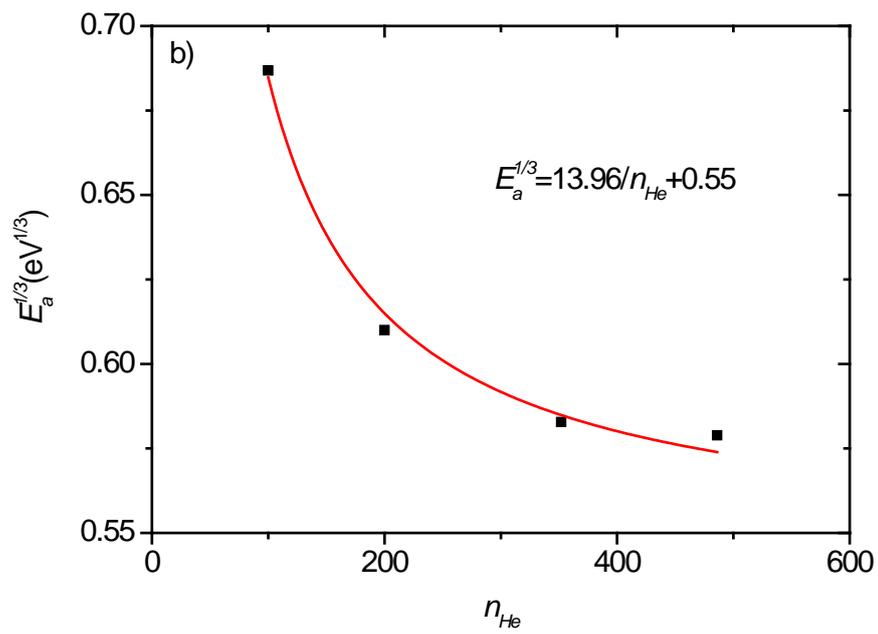

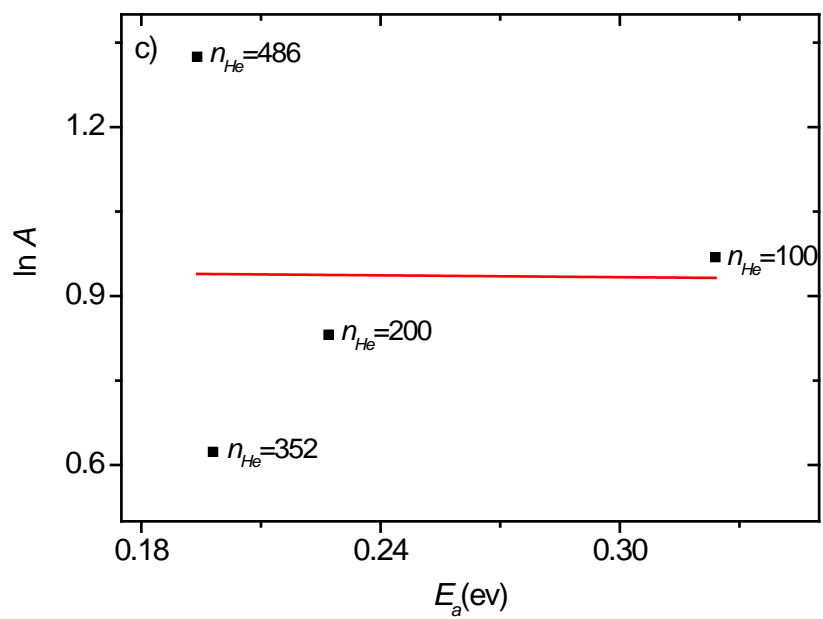

**Fig. 9.** Evolution of bubble pressure vs. temperature for different bubble sizes. For each bubble size, the pressure in the LIT scheme of different $\beta$ values are drawn.

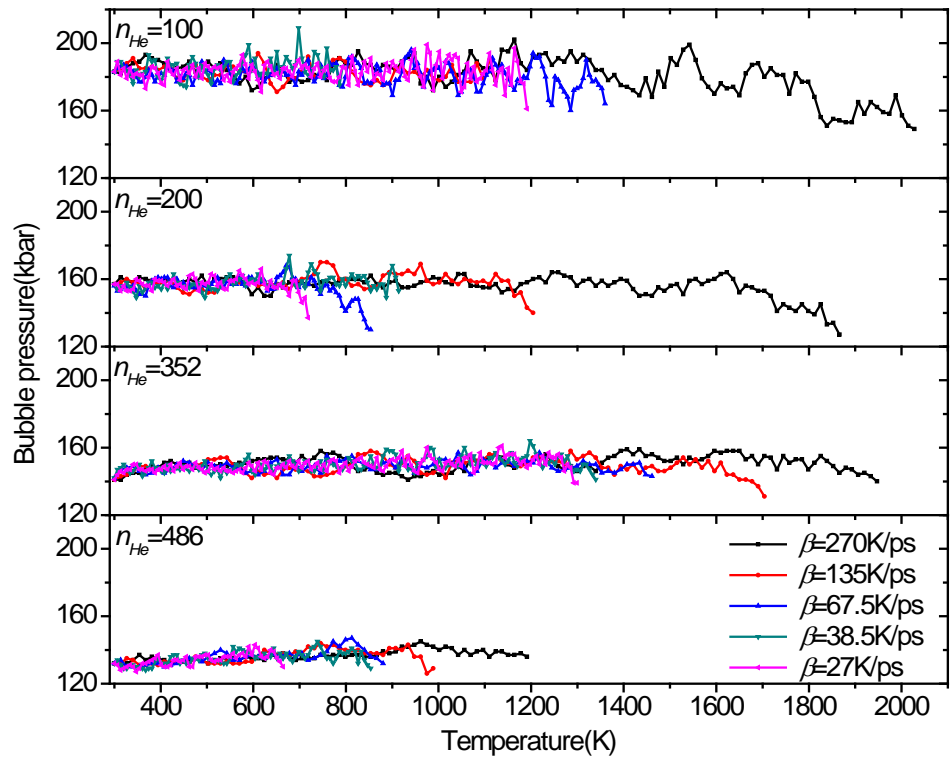